\documentclass[pageno]{jpaper}

\usepackage[normalem]{ulem}
\usepackage{xspace} 

\usepackage[normalem]{ulem}
\usepackage{epsfig}
\usepackage{graphicx}
\usepackage{balance}
\usepackage{comment}
\usepackage{cite}
\usepackage{textcomp}
\usepackage{listings}
\usepackage{color,soul}
\usepackage{colortbl}
\usepackage[dvipsnames]{xcolor}
\usepackage{leading}
\usepackage{moreverb}
\usepackage{listings}
\usepackage{framed}
\usepackage{multirow}

  {\begin{list}{$\bullet$}%
     {\setlength{\parsep}{0pt}%
      \setlength{\topsep}{0pt}%
      \setlength{\itemsep}{2pt}}}%
  {\end{list}}

\newcommand\sname{Glider\xspace}
\newcommand\name{library driver\xspace}
\newcommand\Name{Library driver\xspace}
\newcommand\NAME{Library Driver\xspace}
\newcommand\Names{Library drivers\xspace}
\newcommand\names{library drivers\xspace}
\newcommand\core{device kernel\xspace}
\newcommand\Core{Device Kernel\xspace}
\newcommand\lib{device library\xspace}

\newcommand\libs{device libraries\xspace}
\newcommand\old{legacy\xspace}
\newcommand\Old{Legacy\xspace}

\newcommand\radeont{35}
\newcommand\intelt{38}
\newcommand\radeona{84}
\newcommand\intela{90}

\begin{document}

\title{\sname: A GPU \NAME for Improved System Security}

\author{
{\large
Technical Report 2014-11-14, Rice University
}
\\
\\
{\large 
     Ardalan Amiri Sani, Lin Zhong, Dan S. Wallach
}
\\
{\normalsize
     Rice University
}
}
\date{}
\maketitle

\thispagestyle{empty}

\begin{abstract}

\Old device drivers implement both device resource management and
isolation. This results in a large code base with a wide high-level
interface making the driver vulnerable to security attacks. This is
particularly problematic for increasingly popular accelerators like GPUs 
that have large, complex drivers. We solve this problem with
{\em \names}, a new driver architecture. A \name implements resource
management as an untrusted library in the application process address space, and
implements isolation as a kernel module that is smaller and has a
narrower lower-level interface (i.e., closer to hardware)
than a \old driver. We articulate a set of device and platform
hardware properties that are required to retrofit a \old driver
into a \name. To demonstrate the feasibility and superiority of \names,
we present {\em \sname}, a \name implementation for two GPUs of popular
brands,
Radeon and Intel. \sname\ reduces the TCB size and attack
surface by about \radeont\% and \radeona\% respectively for a Radeon HD 6450 GPU and
by about \intelt\% and \intela\% respectively for an Intel Ivy Bridge GPU. Moreover,
it incurs no performance cost. Indeed, \sname\ outperforms a \old
driver for applications requiring intensive interactions with the device
driver, such as applications using the OpenGL immediate mode API.

\end{abstract}

\section{Introduction}

Device drivers are the main sources of bugs in operating
systems~\cite{Ganapathi2006}. They are large, fast-changing, and
developed by third parties. Since the drivers run in the kernel of
modern monolithic operating systems and are fully shared by
applications, their bugs are attractive targets for exploitation by attackers,
and pose great risks to the security of the system.

This longstanding problem is particularly troubling for hardware
accelerators such as GPUs because they tend to have large, complex
device drivers. For example, GPU device drivers have tens of
thousands of lines of code and have seen quite a few attacks
recently~\cite{BUG4, BUG1, BUG2, BUG3, BUG5}. This problem is increasingly critical as GPUs
are used even by untrusted web applications (through the WebGL
framework). Indeed, browser vendors are aware of the this security risk,
disabling the WebGL framework in the presence of GPU drivers that they
cannot trust~\cite{WebGL_Security}. Obviously, this is a rough solution
and provides no guarantees even in the presence of other drivers.

Researchers have long attempted to protect the
system from device drivers. For example, user space device drivers are
one of the principles of microkernels~\cite{Elphinstone2013}. Even for
monolithic operating systems, there exist solutions that move the driver to 
user space~\cite{Ganapathy2008}, move it to a VM~\cite{Levasseur2004},
or even sandbox it in-situ~\cite{Swift2003, Zhou2006}. Unfortunately,
these solutions have yet to see any practical success, mainly due to their
inferior performance.

In this work, we revisit this problem with a fresh insight: a large part
of \old drivers is devoted to device resource management. Our
solution, inspired by library operating systems, such as
Exokernel~\cite{Engler1995} and Drawbridge~\cite{Porter2011}, is a {\em
\name} design, built on one fundamental principle: {\em
untrusted resource management}.

The \name design incorporates this principle in two steps.
First, it separates device resource management code from resource
isolation. In a \name design, resource isolation is implemented
in a trusted {\em \core} in the operating system kernel, and resource
management is pushed out to the user space. Second, resource management
is implemented as an untrusted library, i.e., a {\em \lib}.
That is, each application that intends to use the device loads and uses
its own \lib. Based on some scheduling policy, the \core\
exports the device hardware resources securely to applications, which
manage and use the resources with their own \lib.

The \name design improves overall system security by reducing the size and
attack surface of the Trusted Computing Base (TCB). With a \old driver,
the whole driver is part of the TCB. However, with a \name,
only the \core, which is smaller than a \old driver, is part of the
TCB. Moreover, compared to a \old driver, the \core\ exposes a narrower
lower-level interface to untrusted software, hence reducing the attack
surface of the TCB. The security benefits of a \name are two-fold:
first, a \name reduces the possibility of attacks on the operating
system kernel through bugs in the driver. Second, it improves the isolation
between applications using the device, as it reduces the amount of shared
state between them. Importantly, a \name improves the system security
without hurting the performance. Indeed, a \name can even outperform
a \old driver due to one fundamental reason: a \name avoids the overhead
of syscalls and user-kernel data copies since it is in the same address
space and trust domain as the application. The performance improvement
highly depends on the application; the more interactions there are
between the application and the driver, the more significant the
performance improvement will be.

Applying the principle of untrusted resource management to a device
driver requires certain hardware properties on the device and platform.
We articulate these properties into three requirements: {\em memory
isolation primitives} such as an IOMMU, {\em innocuous device management
interface}, and {\em attributable interrupts}. If a device and its
platform meet these properties, then it is possible to retrofit the
device resource management code in the form of an untrusted library.
Every violation of these requirements, however, forces some of the
resource management code to remain in the \core, resulting in weaker
security guarantees. 

We target \names mainly for accelerators, such as GPUs, for three
reasons. First, they are an increasingly important subset of devices;
we anticipate various accelerators to emerge in the near future. Second,
they are sophisticated devices requiring large device drivers, in
contrast to simpler devices such as a mouse. Third, they often meet the
hardware requirements mentioned above, as we will discuss in
\S\ref{sec:hardware}.

Based on the aforementioned principle, we implement \sname, a Linux
\name implementation for two GPUs of popular brands, namely the Radeon
HD 6450 and Intel Ivy Bridge GPUs. We implement \sname based on the
original \old Linux drivers. The \name design allows us to implement
both the \core and the \lib by retrofitting the \old driver code, which
significantly reduces the engineering effort compared to developing them
from scratch. We present a full implementation for the Radeon GPU and
a proof-of-concept implementation for the Intel GPU. 

Our evaluation shows that \sname improves the security of the system
by reducing the size and attack surface of the TCB by about \radeont\%
and \radeona\% respectively for the Radeon GPU and by about \intelt\% and \intela\%
respectively for the Intel GPU. We also show that \sname\ provides at
least competitive performance with a \old driver, while slightly
outperforming it for applications requiring intensive interactions with
the driver, such as GPGPU applications with a small compute kernel and
graphics applications using OpenGL's immediate mode.

Beyond monolithic operating systems, \names can benefit other
systems as well. For example, they can be integrated into a library operating
system, such as Exokernel~\cite{Engler1995} or 
Drawbridge~\cite{Porter2011}, or into sandboxing solutions such as
Embassies~\cite{Howell2013}, Xax~\cite{Douceur2008}, and Bromium
micro-virtualization~\cite{bromium_white_paper}, to securely
support multiplexed access to devices.

\section{Library Driver: Design \& Requirements}

In this section, we discuss the \name design, and elaborate on
hardware properties necessary to implement it.

\subsection{Design}
\label{sec:design}

The \name design is based on an important principle from the library
operating system research~\cite{Engler1995}: resource management shall
be implemented as untrusted libraries in applications. Resource management
refers to the code
that is needed to program and use the device. In contrast, \old
device drivers implement resource management along with device resource
isolation in the kernel, resulting in a large TCB with a wide
high-level interface. 

\begin{figure}[t]
\centering
\subfloat[]{
\includegraphics[width=0.49\columnwidth]{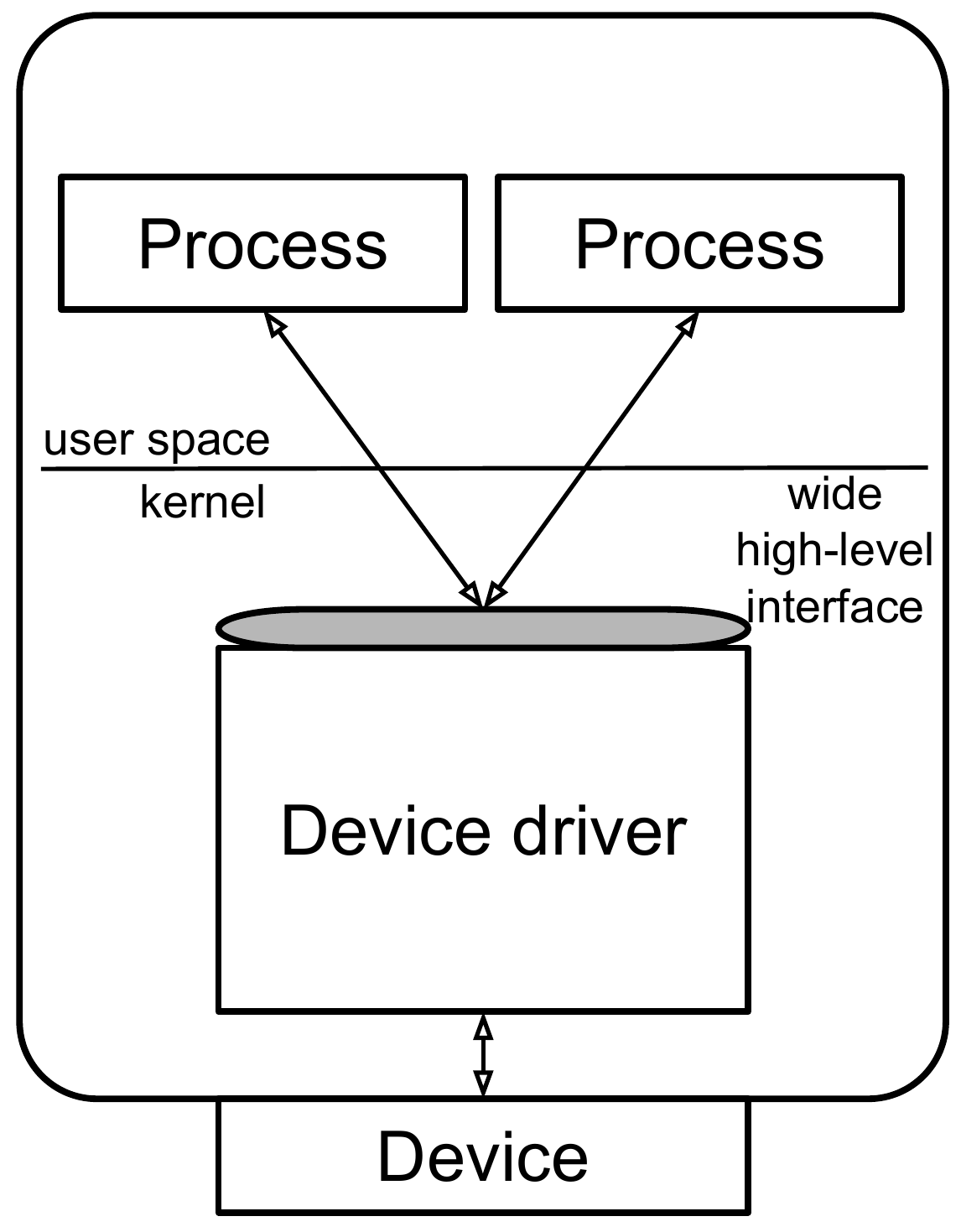}
}
\subfloat[]{
\centering
\includegraphics[width=0.49\columnwidth]{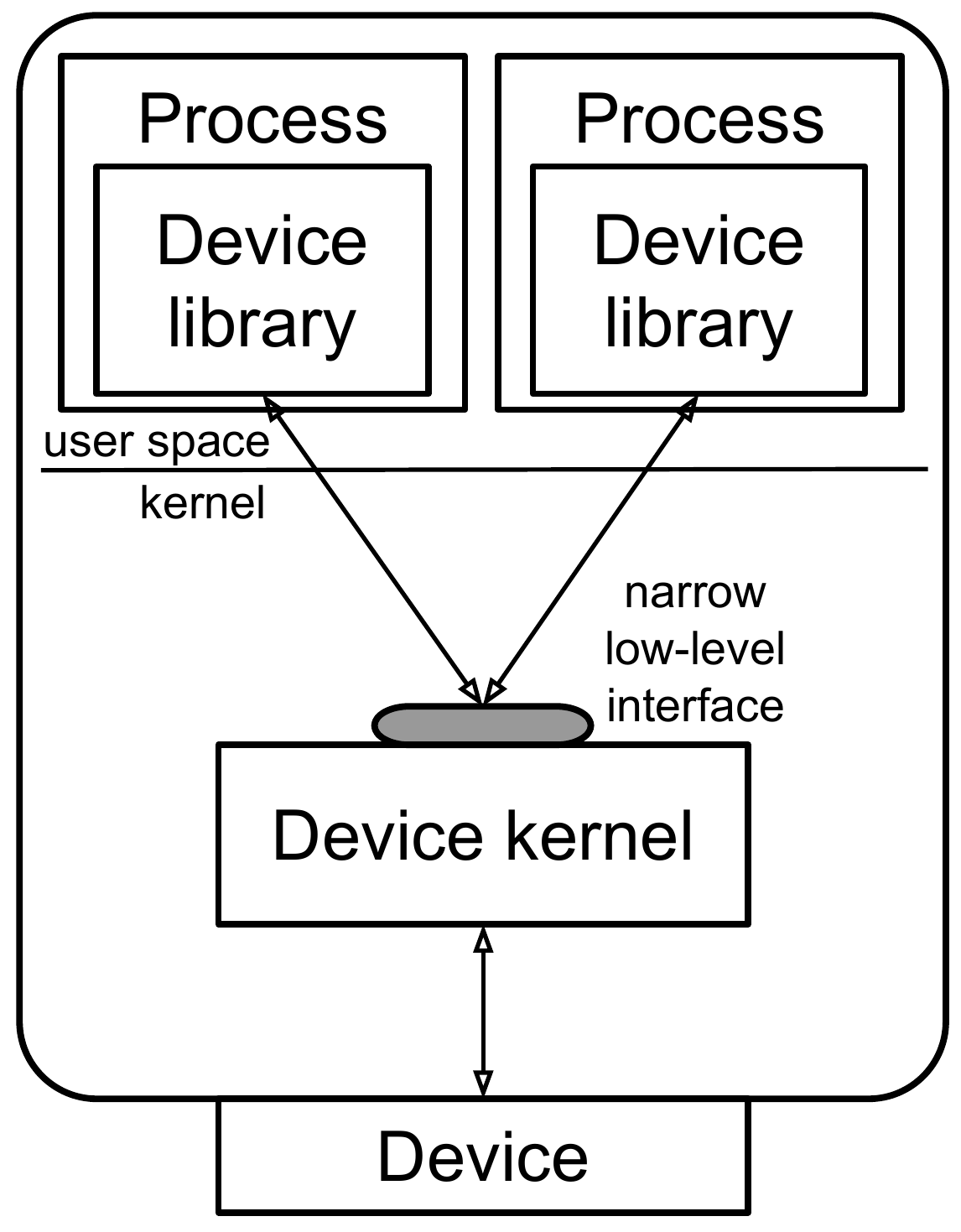}
}
\caption{(a) \Old driver design. (b) \Name design.}
\label{fig:design}
\end{figure}

Figure~\ref{fig:design} compares a \name against a \old driver. 
The \name design applies the aforementioned principle in two steps. First, it separates resource
management from resource isolation. It enforces resource isolation in
a trusted {\em \core} in the operating system kernel and pushes resource management into
user space. Second, the \name design retrofits the resource
management code into an untrusted library, i.e., a {\em \lib}, which is
loaded and used by each application that needs to use the device.

The \core\ is responsible for securely multiplexing the device resources
between untrusting \libs. Based on some scheduling policy, it {\em
binds} the device resources to a \lib at the beginning of a scheduling
epoch and {\em revokes} them at the end of the epoch. If possible, the
\core\ preempts the execution upon revoke. When hardware preemption is
impossible or difficult to implement, e.g., GPUs~\cite{Menon2012}, the
\core\ revokes access when allowed by the device\footnote{It is up to
the scheduling policy to ensure fairness, using techniques similar to
the ones used in~\cite{Menychtas2014, Menychtas2013, Kato2011,
Gupta2011}}. Many devices, including the ones in our prototype, support
only one hardware execution context. For them, the \core\ effectively
time-multiplexes the device between \libs. These devices are the main
focus of this paper. Some devices, however, support multiple hardware
execution contexts~\cite{nvidia_white_paper, Raj2007, Willmann2007,
Dong2008, VMDq}. For those devices, the \core\ can dedicate a hardware
execution context to a \lib as will be discussed in
\S\ref{sec:multi_context}.

The \lib is responsible for managing the resources exported to it by the
\core. It implements all the software abstractions and API calls needed
by the application to use the device. For example, a GPU \lib
implements memory management API calls, such as {\tt create\_buffer},
{\tt read\_buffer}, {\tt map\_buffer}, and {\tt move\_buffer}, on top of
the device memory exported to it by the \core. It is important to note
that the \lib implements the software abstraction and APIs in user
space whilst \old device drivers implement everything in the kernel. We also
note that today's devices often employ user space libraries to create
a unified interface to different drivers of the same device class, e.g.,
GPUs. Such user space libraries do not qualify as \libs since they do
not implement resource management.

\subsubsection*{Key Benefits}
The \name design improves system security because it reduces the
size and attack surface of the TCB. In a \name, only the \core is part
of the TCB. Moreover, without any management responsibilities, the
\core\ is able to expose a narrower lower-level interface to
untrusted software. As a result, the \name design improves two aspects
of system security: it raises the bar for attacks that exploit
device driver bugs, and it improves the isolation between applications
by reducing the amount of shared state between them.

The \name design improves the system security without hurting the
performance. Indeed, a \name can even outperform \old drivers. This is
because a \name eliminates the overhead of the application
interacting with the driver in the kernel, including costly
syscalls~\cite{Soares2010} and user-kernel data copies. In a \name, most
of the interactions occur inside the process in the form of function calls
rather than syscalls. Moreover, because the \lib is in the same trust
domain as the application, data can be passed by reference rather than
being copied, which is commonly done in a \old driver.

As we will demonstrate, the \name design allows us to implement both
the \core and the \lib by retrofitting the
\old driver code. This significantly reduces the engineering
effort to develop \names for existing devices compared to developing
them from scratch.

In addition to security and performance, a \name has other advantages
and disadvantages. In short, the advantages include the possibility of
driver customization for applications, easier user space development and
debugging, and improved operating system memory usage accounting. The
disadvantages include difficulty of multi-process programming,
application launch time overhead, and coarser-grained device memory
sharing for some devices. \S\ref{sec:pros_cons} discusses these issues
in greater detail.

We find that \names are particularly useful for accelerators such as GPUs, 
which are increasingly important for a wide range of applications, from data centers
to mobile devices. Accelerators are sophisticated hardware components and usually come with 
large, complex device drivers, particularly prone to the wide variety of driver-based security exploits. 
More importantly, we have found that accelerators like GPUs often have the necessary hardware
properties to implement a \name, elaborated next.

\subsection{Hardware Requirements}
\label{sec:hardware}

Despite its benefits outlined above, \names cannot support all devices.
The system must have three hardware properties for the \core
to enforce resource isolation. The lack of any of these properties will
leave certain resource management code in the trusted \core.
(\textit{i}) First, in order for each \lib to securely use the memory
exported to it by the \core, hardware primitives must be available to
protect the part of the system and device memory allocated for a \lib
from access by other \libs. (\textit{ii}) Second, in order for a \lib
to safely program the device directly from user space, the registers and
instruction set used to program and use the device must not be
{\em sensitive}, i.e., they must not affect the isolation of resources.
(\textit{iii}) Finally, in order for the \core\ to properly forward the
interrupts to the \libs without complex software logic, the device
interrupts must be easily attributable to different \libs.
Apparently not all
devices have these properties;
fortunately, accelerators
of interest to us, e.g., GPUs, do have these properties, as elaborated below. 

\subsubsection{Memory Isolation Primitives}
\label{sec:hw_req_mem}

The \name design requires hardware primitives to isolate the access to
memory by different \libs. There are two types of memory: system memory
and device memory, the latter being memory on the device itself.
Isolation needs to be enforced for two types of
memory access: \emph{system-side} access via the CPU's load and store
instructions and \emph{device-side} access via direct device programming. 
Modern processors readily provide protection for system-side
access with their memory management units (MMU). As a result, we only
need to be concerned with protecting device-side accesses.

{\bf Device-side access to system memory}:~~The \lib can program the
device to access system memory via direct memory access (DMA).
Therefore, the \core must guarantee that the device only has DMA access
permission to system memory pages allocated (by the operating
system kernel) for the \lib currently using the device. The I/O memory
management unit (IOMMU) readily provides this protection by translating
DMA target addresses. The \core\ can program the IOMMU so that any DMA
requests from a \lib are restricted to memory previously allocated for
it. We observe that IOMMUs are available on most modern systems based on
common architectures, such as x86 and ARM ~\cite{Abramson2006,
Cortex_A15_TRM}. Moreover, GPUs typically have a built-in IOMMU, which
can be used by the \core\ even if the platform did not have an IOMMU. 

{\bf Device-side access to device memory}:~~The \core allocates the
device memory for different \libs, and it must protect them against
unauthorized access. A \lib can program a device to access the device
memory, and such an access does not go through the IOMMU. Therefore, the device must
provide hardware primitives to protect memory allocated for one \lib
from access by another.
There are different forms of memory protection that a device
can adopt, e.g., segmentation and paging. Each form of
memory protection has its pros and cons. For example, segmentation is
less flexible than paging since the allocations of
physical memory must be contiguous. On the other hand, paging is more
expensive to implement on a device~\cite{Pichai2014}.

Isolating access to the device memory only applies to devices that
come with their own memory. We note that accelerators packing their own
memory, such as discrete GPUs and DSPs, often support some form of memory
protection primitives. For example,
NVIDIA GPUs (nv50 chipsets and
later) support paging~\cite{nvidia_hw_doc}
and TI Keystone DSPs support segmentation~\cite{TI_keystone}.

A \old driver does not require such memory protection primitives as it
can implement software-based protection. A \old driver implements the
memory management code in the kernel and employs runtime
software checks to ensure that untrusted applications never program the device to
access parts of the memory that have not been allocated for them.

\subsubsection{Innocuous Device Management Interface}
\label{sec:hw_req_reg}

The \name design further requires the device management interface to be
\textit{innocuous} or not \textit{sensitive}, as defined by the
Popek-Goldberg theorem about virtualiziblity of an
architecture~\cite{Popek1974}. According to Popek and Goldberg,
sensitive instructions are those that can affect the isolation between
virtual machines.

A device usually provides an interface for software to use it.
This interface consists of registers and potentially an instruction set to be
executed by the device. The device management interface is the 
part of the interface that is used for
resource management. Examples include the GPU interface used to dispatch
instructions for execution and the GPU interface used to set up DMA
transfers. Other parts of the programming interface are used for initializing the
device and also to enforce isolation between resources. Examples are the
GPU interface used to load the firmware, the interface used to
initialize the display connectors on the board, the interface used to
prepare the device memory and memory protection primitives.

For {\bf registers}, this requirement means that
management registers cannot be sensitive. That is, registers needed for
resource management cannot affect the resource isolation. For example
on a GPU, software dispatches instructions to the GPU for execution by
writing to a register. This register is part of the management
interface, and therefore must not affect the isolation, e.g., change the
memory partition bounds.

For the {\bf instruction set}, this requirement means
that either the instruction set has no sensitive instructions, or
the sensitive instructions fail when
programmed on the device by untrusted code, such as a \lib.
The latter is more expensive to implement in the device as it requires
support for different privilege levels, similar to x86 protection rings.

Commodity accelerators usually meet this requirement. This is because
registers often have simple functionalities, hence management registers
are not used for sensitive tasks.
(\S\ref{sec:gpu_display} discusses one violation of this requirement).
Also, the instruction set often does not contain sensitive instructions,
and resource initialization and isolation is done through registers
only. 

A \old driver does not need the device to meet the innocuous device
management interface requirement. First, it does not allow untrusted
software to directly access registers. Second, all the instructions
generated by the application are first submitted to the driver, which
can then perform software checks on them to guarantee that sensitive
instructions, if any, are not dispatched to the device.

\subsubsection{Attributable Interrupts}

A \lib using the device must receive the device
interrupts in order to properly program and use the device. For 
example, certain GPU interrupts indicate that the GPU has finished executing
a set of instructions. The interrupt is first delivered to
the \core, and therefore, the \core must be able to
redirect interrupts to the right \lib without the need for any complex
software logic.

This requirement is simply met for commodity accelerators with a single
hardware execution context, since all interrupts belong to the \lib
using the device. On the other hand, we note that this
requirement can be more difficult to meet for non-accelerator devices.
One example is network interface cards with a single receive queue.
Upon receiving a packet (and hence an interrupt), the \core\ cannot
decide at a low level which \lib the packet belongs to. This
will force the \core\ to employ some management code, e.g., packet
filters, to redirect the interrupts, similar to the solutions adopted by
the Exokernel~\cite{Engler1995} and U-Net~\cite{Von1995}.

A \old driver does not have this requirement because it incorporates all
the management code that uses the interrupts. Device events are then
delivered to the application using higher-level API.

\section{GPU Background}
\label{sec:background}

\begin{figure}[t]
\centering
\includegraphics[width=0.75\columnwidth]{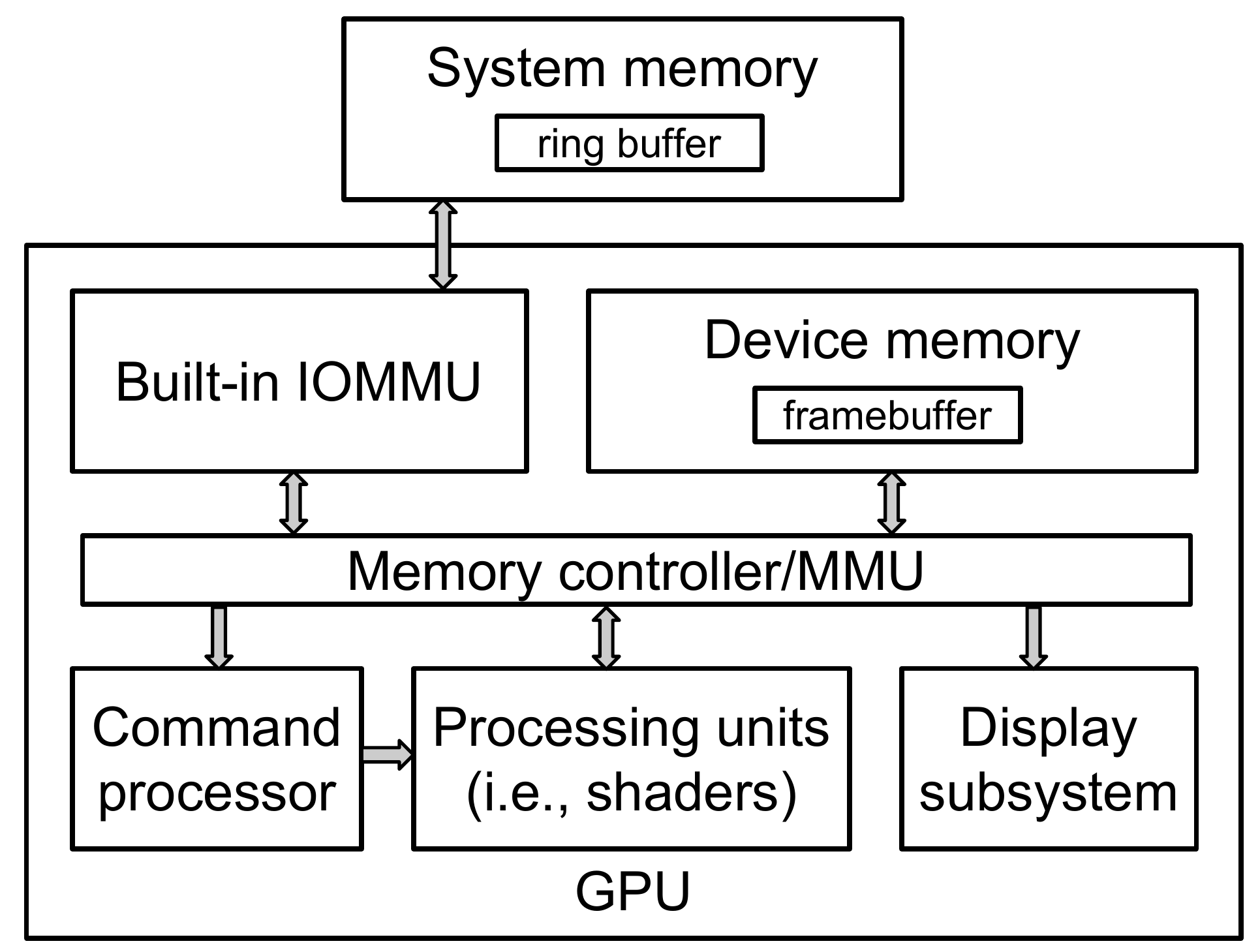}
\caption{Simplified GPU hardware model.}
\label{fig:gpu}
\end{figure}

Before presenting our \name design for GPUs in
\S\ref{sec:glider}, we provide background on the functions of a GPU
device driver and GPU hardware as illustrated in Figure~\ref{fig:gpu}.
A GPU driver has three functions: hardware initialization, resource
management for applications, and resource isolation.

\subsubsection*{Initialization}
GPU hardware initialization loads the
firmware, sets up the GPU's memory controller or MMU in order to
configure the address space used by the GPU, enables the
interrupts, sets up the command processor (which is later used to dispatch instructions to
the GPU), the device power management, and the display subsystem.

\subsubsection*{Management}
Once the GPU hardware is initialized, the driver performs resource
management for applications. It implements four important
functionalities for this. 
First, it implements memory management API
calls. An application can request three types of memory buffers: buffers
on the device memory, buffers on the system memory, and buffers on the
system memory accessible to the GPU for DMA. For the latter, the driver
allocates a range of GPU addresses for the buffer and programs the
GPU's built-in IOMMU to translate from these addresses to their actual
physical addresses on the system memory. Moreover, the driver pins
these pages so that they do not get swapped out by the operating system
kernel.

Second, the driver accepts GPU instructions from applications and
dispatches them to the GPU for execution. Dispatching the instructions
in done through a ring buffer. The driver writes the instructions onto
the ring buffer, and the command processor unit on the GPU reads them and
forwards them to the processing units, i.e., shaders, in the GPU. Note that similar
to CPUs, GPUs cache memory accesses and the built-in IOMMU has a TLB for
its translations. Therefore, the driver flushes the caches and the TLB
as needed.

Third, the driver handles interrupts from the GPU. Most importantly, it
processes the interrupts that indicate the end of the execution of the
instructions by the processing units. The driver then informs the application
so that the application can send in new instructions to be executed.

Finally, the driver implements some API calls for the application to
allocate and use a framebuffer and to set the display mode. The
framebuffer is a memory buffer holding the content that will be
displayed. Through programming the GPU or by directly accessing
the memory, the application can write to the framebuffer. The
display mode includes the resolution, color depth, aspect ratio, etc.
Different applications require to set different modes for the display.
Traditionally, the display mode was set through the X server in Linux.
However, it was recently moved to the \old Linux open source drivers and is
referred to as Kernel Mode Setting, or KMS. The advantage of KMS is that it
allows for better graphics at boot time (before X is even initialized)
and during Linux virtual console switching.

\subsubsection*{Isolation} 
The driver also enforces isolation between applications by performing
appropriate hardware configurations, such as setting up the MMU page tables
if supported, or through software checks. For instance, a \old GPU driver
enforces software memory isolation as follows. When allocating memory
buffers for an application, the driver only returns an ID of the
buffer to the application, which then uses that ID in the instructions
that it submits to the driver. The driver then replaces all the IDs with
the actual addresses of the buffers and then writes them to the ring
buffer.

\vspace{+1ex}
\sname implements these three functionalities as follows.
Initialization is performed in the \core. Once the GPU is initialized,
the \core exports the resources to \libs so that they can perform
resource management in the user space. For isolation, the \core
leverages hardware properties introduced in \S\ref{sec:hardware}. This
design reduces the size and attack surface of the TCB and hence improves
the system security. The GPU hardware often meets the required hardware
properties. It provides either a memory controller or an MMU that can be
used to enforce isolation for device-side accesses to its memory. The
management registers are often not sensitive and the instruction set has
no sensitive instructions. Moreover, with GPUs with a single hardware
execution context, interrupts can be simply forwarded to the \lib using
the GPU.

\section{\sname: \NAME for GPUs}
\label{sec:glider}

In this section, we present \sname,
\names for the Radeon HD 6450 and Intel Ivy Bridge GPUs based on
retrofitting their corresponding \old drivers. We provide
a fully-functional implementation for the Radeon GPU and
a proof-of-concept implementation for the Intel GPU.

\subsection{Isolation of GPU Resources}
\label{sec:gpu_architecture}

We first identify the important GPU hardware resources and elaborate on
how the \core\ securely exports them to \libs.

\subsubsection{Processing Units}

The \core securely exports the GPU processing units, i.e., shaders, by
allowing access to the GPU command processor. When binding the GPU to
a \lib, the \core allows the \lib to update the command processor's
registers that determine the ring buffer location. This enables the \lib
to allocate and use the ring buffer that it has allocated from its own
memory. The \lib then populates the ring buffer with instructions from
the application and triggers the execution by writing to another
register. At revoke time, the \core\ disallows further write to the
trigger registers. It then waits for ongoing execution to finish on the
GPU before binding the GPU to another \lib. The \core also takes
a snapshot of the registers updated by the \lib so that it can write
them back the next time it needs to bind the GPU to the same \lib.
Finally, it resets the command processor and flushes the GPU caches
appropriately.

\subsubsection{Memory}
\label{sec:gpu_memory}

The operating system allocates system memory pages for a \lib through standard
syscalls, such as {\tt mmap} and the \core allocates device memory for
the \lib through its API calls (\S\ref{sec:apis}). When using the device,
the \core needs to guarantee that a \lib can only access memory
allocated for it, as discussed in \S\ref{sec:hw_req_mem}. Here, we
provide the implementation details of isolating device-side access to
memory.

For system memory, we use the system IOMMU for isolation. The \core\
provides an API call for the \lib to map a page into the IOMMU. More
specifically, the \lib can ask the \core to insert a mapping into the
IOMMU in order to translate a DMA target address to a given physical
page address. To enforce isolation, the \core\ only maps pages that have
been allocated for the \lib. The \core\ also pins the page into memory
so that it does not get swapped out or deallocated by the operating
system kernel. Upon revoke, the \core\ stores all the mappings in the
IOMMU and replaces them with the mappings for the next \lib. The
mappings of an IOMMU is in the form of page tables~\cite{Abramson2006},
very similar to the page tables used by the CPU MMU. Therefore, similar
to a context switch on the CPU, changing the IOMMU mappings can be done
very efficiently by only changing the root of the IOMMU page tables.

Alternatively, the GPU's built-in IOMMU can be used for isolation, which
is useful for systems
without a system IOMMU. To demonstrate this, we used the built-in IOMMU
for our Intel GPU \name. In this case, the \core\ takes full control of the GPU's
built-in IOMMU and updates its page tables through the same API call
mentioned above.

It is up to the \lib when and how much memory it maps in the IOMMU. In
our current implementation, the \lib allocates about 20 MB of memory at
launch time and maps all the pages in
the IOMMU. We empirically determined this number to be adequate for our
benchmarks. Alternatively, a \lib can allocate and map the pages in the
IOMMU as needed. This alternative option speeds up the \lib's launch
process but may result in degraded performance at runtime.
 
For the Radeon device memory, we use the GPU memory controller for
isolation. Isolating the device memory does not apply to the Intel GPU
since it does not have its own memory. The memory controller on the
Radeon GPU configures the physical address space seen by the GPU.
It sets the range of physical addresses that are forwarded to the GPU
memory and those that are forwarded to the system memory (after
translation by the built-in IOMMU). By programming the memory
controller, the \core can effectively segment the GPU memory, one
segment per \lib. Obviously, the memory segmentation is not flexible in
that each \lib can only use a single contiguous part of the GPU memory.
Changing the segments can only be done using memory ballooning
techniques used for memory management for VMs~\cite{Waldspurger2002},
although we have not implemented this yet.

\subsubsection{Framebuffer and Displays}
\label{sec:gpu_display}

The hardware interface for the framebuffer can be securely exported to
\libs. The \core\
allows the \lib to have access to the registers that determine
the location of the framebuffer in memory. Therefore, the \lib
can allocate and use its own framebuffer. However, the hardware
interface for display mode setting violates the requirement in
\S\ref{sec:hw_req_reg}, forcing us to keep the display mode setting code
in the \core.

Every display supports a limited number of modes that it can
support. However, instead of exposing the mode option with a simple
interface, e.g., a single register, GPUs expect the software to
configure voltages, frequencies, and clocks of different components on
the GPU, e.g., display connectors and encoders, to
set a mode, and it is not clear whether such a large hardware interface
can be safely exported to unstrusted software without risking damaging
the device. Exacerbating the problem, newer Radeon GPUs have
adopted mode setting through GPU BIOS calls. However, BIOS calls can
also be used for other sensitive GPU configurations, such as for power
management and operating thermal controllers, and we cannot export the BIOS
register interface to \libs securely.

As a result, we keep the display mode setting code in the \core. The
\core\ exposes this mode setting functionality with a higher level API
call to \libs. This results in a larger TCB size as is reported in
\S\ref{sec:eval_security}. Despite its disadvantage of increasing the
TCB size, keeping the mode setting in the \core\ has the advantage that
it supports Kernel Mode Setting (\S\ref{sec:background}).

\subsection{The \Core API}
\label{sec:apis}

The \core API include seven calls for \libs and two calls
for the system scheduler. Except for the {\tt set\_mode} call, which
is GPU-specific (\S\ref{sec:gpu_display}), the
rest are generic. Therefore, we expect them to be used for \libs of
other devices as well. They constitute the minimal set of API calls to
support \lib's secure access to system memory, device memory,
and registers. The seven calls for \libs are as follows:

{\tt void *init\_device\_lib(void)}: This call is used when a \lib is
first loaded. The \core\ prepares some read-only memory pages
containing the information that the \lib needs, such as the
display setup, maps them into the \lib's process address space, and
returns the address of the mapped pages to the \lib.

{\tt iommu\_map\_page(vaddr, iaddr)},
{\tt iommu\_unmap\_page(vaddr)}: With these two calls, the \lib asks the
\core\ to map and unmap a memory page to and from the IOMMU. {\tt vaddr}
is the virtual address of the page in the process address space. The
\core\ uses this address to find the physical address of the page. {\tt
iaddr} is the address that needs to to be translated by the IOMMU to the
physical address. This will be the DMA target address issued by the
device.

{\tt alloc\_device\_memory(size)},
{\tt release\_device\_memory(addr, size)}: With these two calls, the
\lib allocates and releases the device memory. These two calls are only
implemented for GPUs with their own memory.

{\tt int access\_register(reg, value, is\_write)}: With
this call, a \lib reads and writes from and to authorized
registers. The implementation of this call in the \core\ is simple: it just
checks whether the read or write is authorized or not, and if yes,
it completes the request. The checking is done by maintaining
a list of authorized registers. We implement read and write
operations in one API call since their implementation is different only
in a few lines of code.

Given that most registers on GPUs are memory-mapped (i.e., MMIO
registers), One
might wonder why the \core does not directly map these registers into
the \lib's process address space, further reducing the attack surface on
the TCB. This is because with such an approach, protection of registers
can be enforced at the granularity of a MMIO page, which contains
hundreds of registers, not all of them are authorized for access by
a \lib. 

{\tt set\_mode(display, mode)}: With this call,
a \lib asks the \core\ to set the mode on a given display.

We next present the two calls for the system scheduler. Schedulers for
GPU resources, such as \cite{Menychtas2014, Menychtas2013, Kato2011,
Gupta2011}, can be implemented on top of these two API calls.

{\tt bind\_device\_lib(id)},
{\tt revoke\_device\_lib(id)}:
With these two calls, the
scheduler asks the \core\ to bind and revoke the GPU resources to and
from a
\lib with a given {\tt id}.
Since our GPUs do not support execution preemption (\S\ref{sec:design}),
the revoke call needs to block until the execution on the GPU terminates.

\subsection{Reusing \Old Driver Code for \sname}
\label{sec:uml}

We reuse the Linux open source \old driver code in \sname, both for the
\core and the \lib, rather than implementing them from the scratch, in order
to reduce the engineering effort. Reusing the \old driver code for \core
is trivial since the \core runs in the kernel as well. However, reusing
it for the \lib is challenging since the \lib is in the user space. We
solve this problem by using the User-Mode Linux
framework~\cite{Dike2001}. UML is originally designed to run a Linux
operating system in the user space of another Linux operating system and
on top of the Linux syscall interface. It therefore provides
translations of kernel symbols to their equivalent syscalls,
enabling us to compile the \lib to run in the user space. We only use
part of the UML code base that is needed to provide the translations for
the kernel symbols used in our drivers.

Note that the UML normally links the compiled object files into an
executable. We, however, link them into a shared library.  Linking into
a shared library may be challenging on some architectures. This is
because assembly code, which is often used in the kernel, is not always
position-independent, a requirement for a shared library. We did not
experience any such problem for the x86 (32 bit) architecture. The
solution to any such potential problem in other architectures is to
either rewrite the assembly code to be position-independent or to
replace it with equivalent non-assembly code.

It is interesting to understand how system memory allocation works in
\sname's \lib, which is compiled against the UML symbols. As mentioned
in \S\ref{sec:gpu_memory}, we allocate about 20 MB of system memory at
the \lib's launch. This memory is then managed by the slab allocator of
UML, similar to how physical memory is managed by the slab allocator in
the kernel. The retrofitted driver code in the \lib then allocates
system memory from the UML's slab allocator by calling the Linux kernel
memory allocation functions, i.e., {\tt kmalloc()} and its derivatives. 

\subsection{Other Implementation Challenges}
\label{sec:gpu_implementation}

We solved two other challenges in \sname. 
First, we replace syscalls for the \old driver with function calls into equivalent entry points in the
\lib. Fortunately in the case of GPU, we managed to achieve this by only
changing about 20 instances of such syscalls in Linux GPU libraries
including the {\tt libdrm}, {\tt xf86-video-ati}, and
GalliumCompute~\cite{GalliumCompute} libraries. An alternative solution
with less engineering effort
is to use the {\tt ptrace} utility to
intercept and forward the syscalls to the \lib. This
solution, however, will have noticeable overhead.

Second, we implement fast interrupt delivery to \libs for good performance.
We experimented with
three options for interrupt delivery. The first two, i.e., using the OS
signals and syscall-based polling, resulted in performance degradation
as these primitives proved to be expensive in Linux. The third option
that we currently use is polling through shared memory. For example, in
the case of the Radeon GPU, in addition to the interrupt, the GPU
updates a memory page with the information about the interrupt that was
triggered, and the \lib can poll on this page. This approach provided
fast interrupt delivery so that the interrupts do not become
a performance bottleneck. However, as we will show in
\S\ref{sec:eval_graphics}, it has the disadvantage of increased CPU
usage.

We are considering two other options that we believe will provide fast
interrupt delivery without the extra CPU usage. The first approach is
using upcalls, which allow the kernel to directly invoke a function in
user space. The second approach is to use the interrupt remapping
hardware available from virtualization hardware extensions. This
hardware unit can deliver the interrupts directly to the \lib's process,
similar to Dune~\cite{Belay12}.

\section{Evaluation}

We evaluate \sname\ and show that it improves the system security
by reducing the size and attack surface of
the TCB. We also show that \sname\ provides at least competitive
performance with a \old kernel driver, while slightly outperforming
it for applications that require extensive interactions with the driver.

\subsection{Security}
\label{sec:eval_security}

We measure the size and attack surface of the TCB, i.e., the whole
driver in the case of a \old driver and the \core in the case of \sname.
Unlike the \old driver that supports various GPUs of the same brand,
\sname\ only supports one specific GPU. Therefore, for a fair
comparison, we remove the driver code for other GPUs as best as we
manually can. This includes the code for other GPU chipsets, other
display connectors and encoders, the code for the legacy user mode
setting framework not usable on newer GPUs (\S\ref{sec:background}), the code for
audio support on GPUs, and the code for kernel features not supported in
\sname, such as with Linux {\tt debugfs}.

Our results, presented in Table~\ref{tab:code}, show that \sname\
reduces the TCB size by about \radeont\% and \intelt\% for the Radeon
and Intel GPUs, respectively. In the same table, we also show the size
of the code in C source files (and not header files). These numbers show
that a large part of the code in \sname\ TCB are headers, which mainly
include constants, such as register layouts. Not including the headers,
\sname\ reduces the TCB size by about 47\% and 43\% for the Radeon and
Intel GPUs.

As discussed in \S\ref{sec:gpu_display}, the display subsystem hardware interface
violates the hardware requirements for a \name resulting
in a larger TCB. To demonstrate this, we measure the code size in
display-related files in the \core. For the Radeon and Intel GPUs, we
measure this number to be 19 and 13 kLoC, which is 50\% and 54\% of the
\sname\ TCB.

We also show the TCB attack surface in Table~\ref{tab:code}. \sname\
reduces the attack surface by \radeona\% and \intela\% for the Radeon
and Intel GPUs. \sname\ only exposes 9 and 7 API calls for these GPUs,
as described in detail in \S\ref{sec:apis} (Intel GPUs do not implement
the two API calls for device memory). In contrast, the \old driver
exposes 56 and 68 API calls for the same GPUs. These large number of
API calls are used for memory management, GPU execution, display mode
setting, and inquiring information about the GPU. \sname\ supports the
first two by securely giving a \lib access to part of the GPU management
interface. It supports mode setting with one API call and supports the
information inquiring API either through the constants compiled into the
\lib or through the read-only information pages mapped into the \lib
(\S\ref{sec:gpu_architecture}).

\subsection{Performance}

In this section, we evaluate the performance of \sname\ for the Radeon
GPU using both compute and graphics benchmarks.

For the experiments, we run the drivers inside a 32-bit x86 (with
Physical Address Extension) Ubuntu 12.04 operating system running Linux
kernel 3.2.0. The machine has a 3rd generation core i7-3770 (with
hyper-threading disabled). We configure the machine with 2 GBs of memory
and 2 cores. In order to minimize the effect of the operating system
scheduler on our experiments, we isolate one of the cores at boot time
using the Linux {\tt isolcpus} command-line boot option. With this
option, Linux only schedules kernel threads on the isolated core, but it
does schedule user application threads on it unless explicitly asked
for. We then pin our benchmarks to the isolated core and set the highest
scheduling priority for them. In order to use the system IOMMU for the
Radeon GPU, we run the benchmarks inside a Xen VM with the same
configurations mentioned above (2 GBs of memory and 2 cores). The Radeon
GPU is assigned to the VM using the direct device
assignment~\cite{Abramson2006, Liu2006, Gordon2012, Ben-Yehuda2010}.

\newcolumntype{C}[1]{>{\centering\let\newline\\\arraybackslash\hspace{0pt}}m{#1}}

\begin{table}[t!]
\small
\centering
\begin{tabular}{|C{0.98cm}|C{1.4cm}|C{1.5cm}|C{1.5cm}|C{1.1cm}|}
\hline
 & & TCB & TCB & API \\
 & & \scriptsize{(all files)} & \scriptsize{(source files)} & calls\\ \hline\hline
\multirow{3}{*}{Radeon}
& \Old  & 55 & 34 & 56 \\ \cline{2-5}
& Glider & 36 & 18 & 9 \\ \cline{2-5}
& Reduction & {\bf 35}\% & {\bf 47}\% & {\bf 84}\% \\ \hline\hline
\multirow{3}{*}{Intel}
& \Old & 39 & 30 & 68 \\ \cline{2-5}
& Glider & 24 & 17 & 7 \\ \cline{2-5}
& Reduction & {\bf 38}\% & {\bf 43}\% & {\bf 90}\% \\ \hline
\end{tabular}
\caption{TCB size and attack surface for the \old and \sname for the Radeon HD
6450 and Intel Ivy Bridge GPUs. The numbers for both TCB columns are
in kLoC. The first TCB columns reports LoC in both source and header
files. The second TCB column excludes the header files.}
\label{tab:code}
\end{table}

\begin{figure}[t]
\centering
\includegraphics[width=0.83\columnwidth]{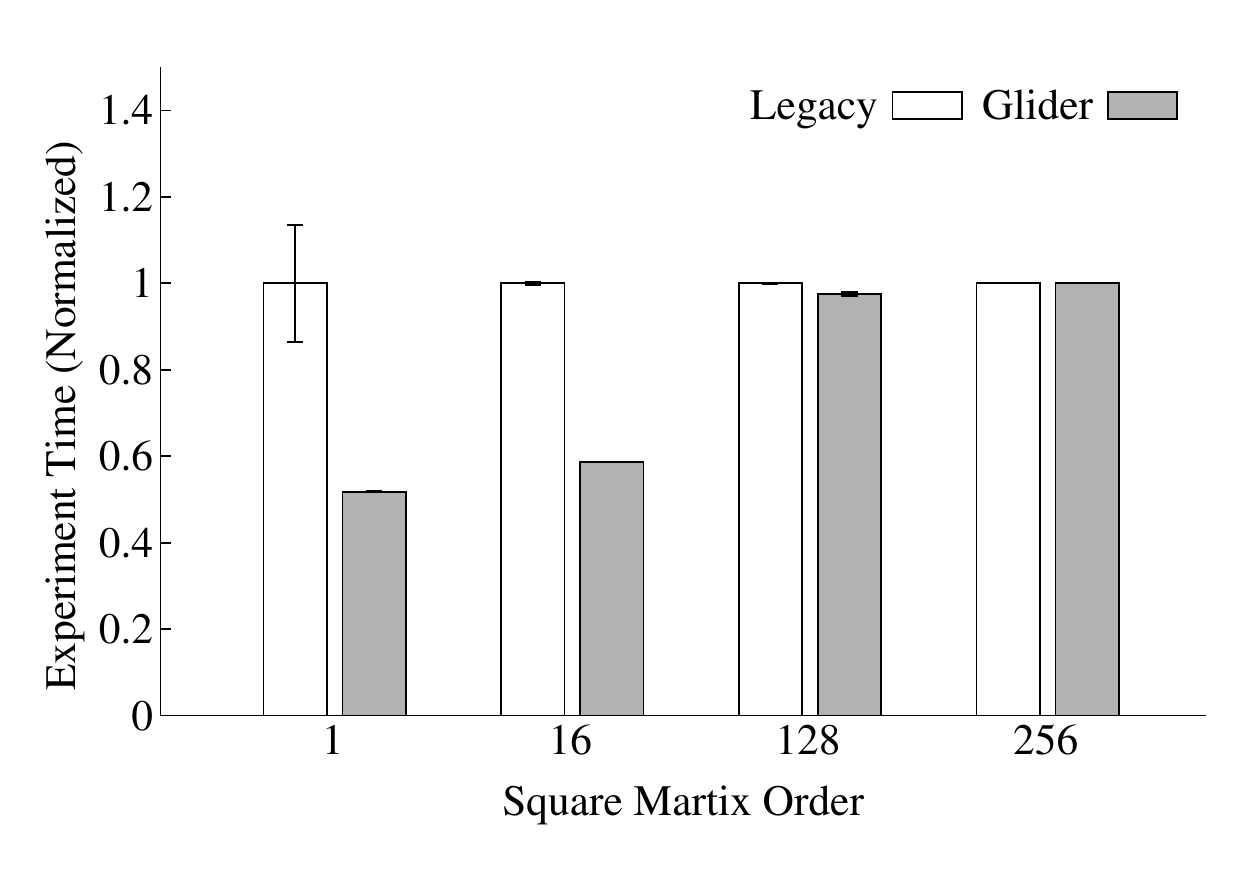}
\caption{OpenCL matrix multiplication. The x-axis
shows the (square) matrix order. The y-axis is the time to execute the
multiplication on the GPU, normalized to the average time by the
\old driver.}
\label{fig:mat_mul}
\end{figure}

\subsubsection{Compute Benchmarks}

We use a matrix multiplication benchmark running on top of the
GalliumCompute~\cite{GalliumCompute} framework, an open source
implementation of OpenCL. We evaluate the performance of multiplying
square matrices of varying orders. For each experiment, we use a host
program that populates input matrices and launches the compute kernel on
the GPU. The programs repeats this for 1000 iterations in a loop and
outputs the average time for a single iteration. We discard the first
iteration to avoid the effect of \sname's launch time overhead (characterized
in \S\ref{sec:eval_overheads}) and we do not include the time to compile
the OpenCL kernel. We then repeat the experiment 5 times for each
matrix size and report the average and standard deviation.

\begin{figure*}[t]
\centering
\begin{minipage}[t]{1\columnwidth}
\centering
\begin{minipage}[b]{0.40\columnwidth}
\centering
\includegraphics[width=1\columnwidth]{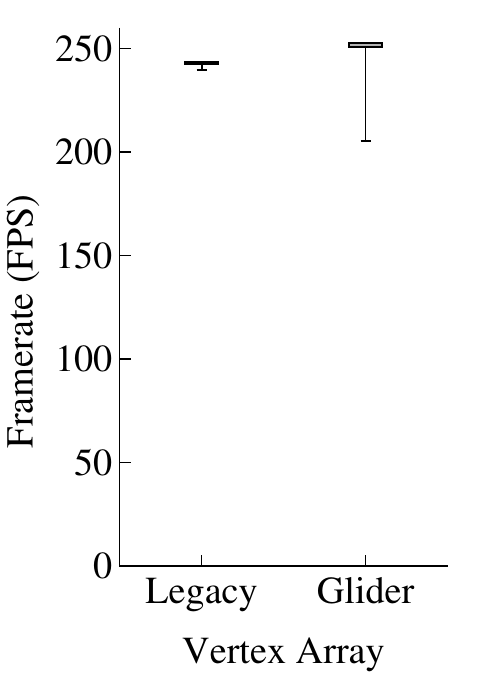}
\end{minipage}
\hspace{0.15in}
\begin{minipage}[b]{0.40\columnwidth}
\centering
\includegraphics[width=1\columnwidth]{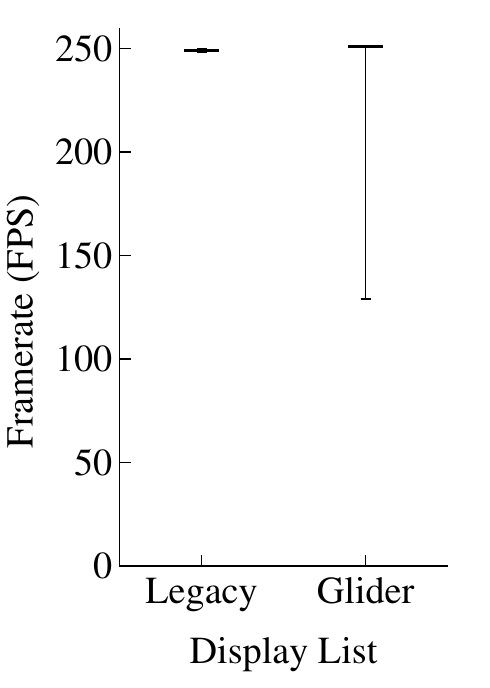}
\end{minipage}
\caption{Graphics performance.
The box plots show 90\% percentile, 10\%
percentile, and median. The whiskerbars depicts maximum and minimum.
The box plots are highly concentrated around the mean.
}
\label{fig:radeon_graphics}
\end{minipage}
\hspace{0.15in}
\begin{minipage}[t]{1\columnwidth}
\centering
\begin{minipage}[t]{0.83\columnwidth}
\centering
\includegraphics[width=1\columnwidth]{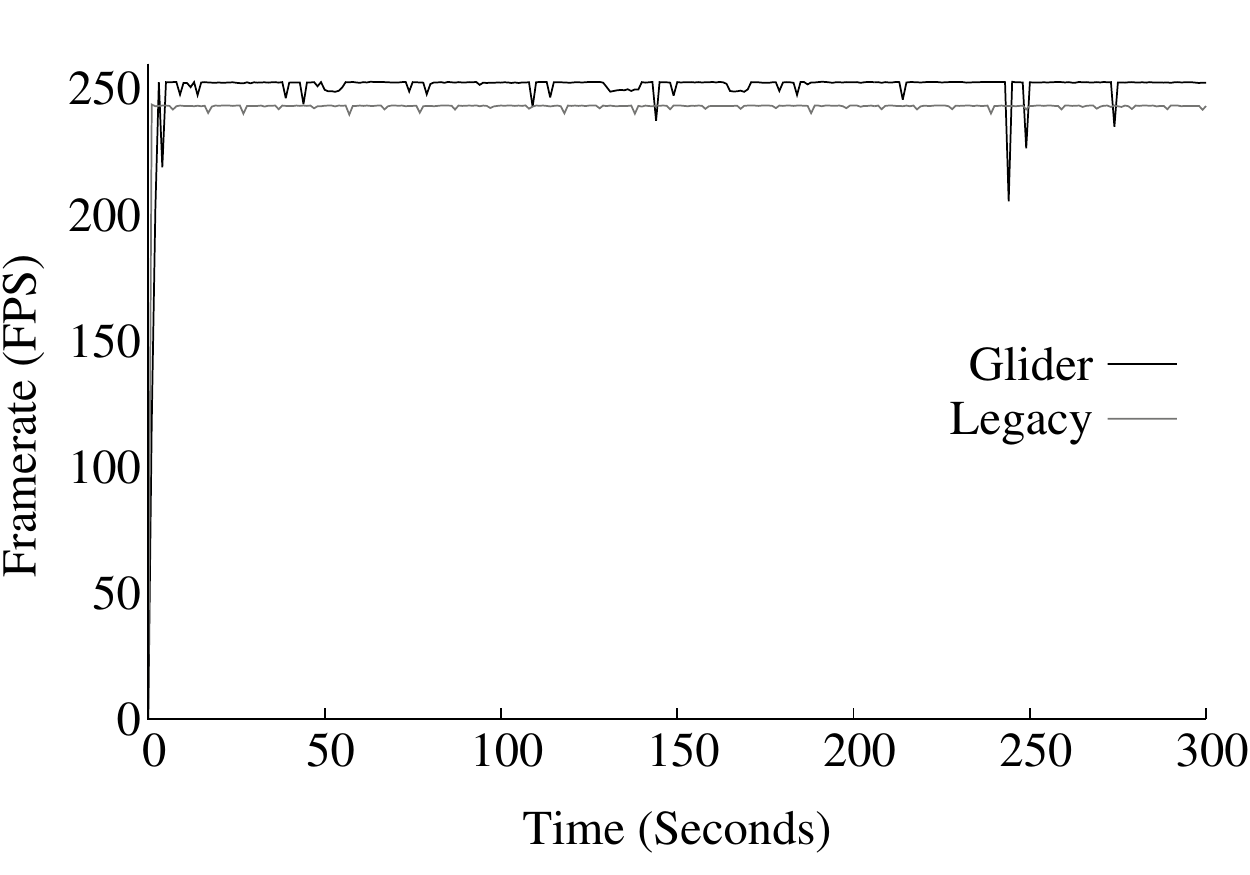}
\end{minipage}
\caption{A sample run of the Vertex Array benchmark. The framerate is
measured every second. The figure shows the rare drops of framerate and
the slow start of \sname.
}
\label{fig:timegraph}
\end{minipage}
\end{figure*}

Figure~\ref{fig:mat_mul} shows the results, normalized to the average
performance by the \old driver in each case.
It shows that \sname\ outperforms the
\old driver for smaller matrix sizes, while providing competitive
performance for all other sizes. This is because for large matrix sizes,
the majority of the time is spent on transferring the data between the system and
device memory and on the GPU executing the kernel. Consequently,
the driver design does not impact the performance noticeably. For
smaller sizes, on the other hand, the overhead of application's interaction
with the driver becomes more significant.

\subsubsection{Graphics Benchmarks}
\label{sec:eval_graphics}

For graphics, we use two OpenGL benchmarks running on top of the MESA
open source implementation of OpenGL~\cite{Mesa}. Both benchmarks draw
a teapot, and update the screen as fast as possible. One benchmark uses
the Vertex Array API~\cite{va} (the immediate mode) and the other uses
the Display List API~\cite{dl}. In each experiment, we run the benchmark
for 5 minutes, recording the framerate every second. We discard the
first 5 frames to avoid the effect of \sname's launch time overhead
(characterized in \S\ref{sec:eval_overheads}). We repeat each
experiment three times.

The results, shown in Figure~\ref{fig:radeon_graphics}, demonstrates
that \sname\ achieves similar performance as the \old driver for the
Display List benchmark but outperforms the \old driver for the Vertex
Array one. In order to explain these results, it is important to
understand these two OpenGL API's. With Vertex Arrays, the program needs
to send the vertices information to the device for every frame, which
requires interacting with the driver quite significantly. On the other
hand, the Display List API allows this information to be cached by the
device in order to avoid resending from the application. Intuitively,
\sname\ improves the performance for the benchmark with more driver
interactions, or the Vertex Array.

We also measure the CPU usage for both drivers when running these
benchmarks. Our results show that the \old driver consumes 53.7\% and
41.7\% of the CPU time for the Vertex Array and Display List benchmarks,
respectively, whereas \sname\ consumes 75.3\% and 71.6\% of the CPU time
for the same benchmarks. The extra CPU usage of \sname\ is due to
polling the memory for interrupts (\S\ref{sec:gpu_implementation}). One
might wonder whether the extra CPU usage is the source of the
performance improvement. To investigate this, we attempted to employ
similar polling methods in the \old driver, but failed to improve the
performance. Therefore, we are convinced that \sname's performance
improvement is due to eliminating the application and driver
interactions' overhead. We also report the performance when using
syscall-based polling in the \lib, which incurs delay in delivering the
interrupt to the \lib. With this method, for the Vertex Array and
Display List benchmarks, respectively, \sname\ consumes 54.3\% and
35.3\% of the CPU time, and achieves 193.3 and 231.5 median frames per
second, which are noticeably lower compared to when the \lib polls the
memory for interrupts.

Figure~\ref{fig:radeon_graphics} also shows that \sname\ achieves
a noticeably lower minimum framerate, although the minimum happens
rarely. To further demonstrate this, we show a sample run of the Vertex
Array benchmark in Figure~\ref{fig:timegraph}. Our investigation shows
that the performance drop is due to OS scheduling despite our attempts
to minimize such effect.

\subsection{Library Driver Overheads}
\label{sec:eval_overheads}

We measure two important overheads of \sname: the \lib's launch
time overhead and the \core's switch time, i.e., the time it takes the
core to revoke the GPU from one \lib and bind it to another.

Figure~\ref{fig:timegraph} illustrates the effect of the \lib's launch
time overhead on the graphics benchmark reported before. It shows that
the performance of \sname\ is inferior to that of the \old driver in the
first few seconds. There are two sources for this overhead: the \lib's
initialization overhead and the UML's slab allocator's initialization
overhead. The former is the time it takes the \lib to initialize itself
so that it is ready to handle requests from the application. We measure
this time to have an average/standard deviation of of 109 ms/7 ms and 66
ms/0.3 ms for the Radeon and Intel GPUs, respectively. The latter is
because the slab allocator of UML (\S\ref{sec:uml}) takes some time to
populate its own cache, very similar to the suboptimal performance of the
slab allocator in the kernel at system boot time.

We also measure the switch time in the \core, as defined above. We measure
this time to have an
average and standard deviation of 42 $\mu$s and 5 $\mu$s for the Radeon
GPU (we have not yet implemented this feature for the Intel GPU). The
switch time consists of the \core taking a snapshot of GPU registers updated by the
current \lib, writing the register values for the new \lib, changing
the IOMMU mappings, resetting the command processor, and flushing the GPU
caches and the built-in IOMMU TLB. These measurements show that changing the
GPU binding can be done fairly quickly.

\subsection{Engineering Effort}

As mentioned, we build \sname by retrofitting the Linux open source drivers as
a baseline. We added about 4 kLoC and 2 kLoC
for the Radeon and Intel \names, respectively. These changes were
applied to 49 and 30 files, and we added two new files in
each case. We have implemented both the \core and the \lib on the same
driver, and the two are differentiated at compile time.
While we
believe that reusing the \old driver code significantly reduced our
engineering effort compared to developing from scratch, we note that
implementing the \names still required noticeable effort. The main source of
difficulty was gaining a deep understanding of the GPU internals and its
device driver with 10s thousands of lines of code. Fortunately, our experience
with the Radeon GPU \name made it easier for us to prototype
the Intel GPU \name. We therefore believe that experienced
driver developers can develop \names without prohibitive engineering effort.

\section{Related Work}

\subsubsection*{Library Operating Systems and Sandboxes}

Library operating systems, such as Exokernel~\cite{Engler1995} and
Drawbridge~\cite{Porter2011} improve the system security by executing
the operating system management components as a library in the
application's process address space. Library drivers are complementary;
they can serve as a secure way for applications to use the devices.

Hardware sandboxing techniques, such as Embassies~\cite{Howell2013},
Xax~\cite{Douceur2008}, and the Bromium
micro-virtualization~\cite{bromium_white_paper}, improve the system
security by reducing the TCB size. Library drivers can complement these
sandboxes by providing them with secure access to devices.

\subsubsection*{Alternative Device Driver Designs}

Many previous solutions have attempted to reduce the drivers' risk on
the system security.
Some solutions move the driver to user space or to a VM. For
examples, in microkernels, drivers reside completely in user space
in order to keep the kernel minimal~\cite{Elphinstone2013, Leslie2005,
Forin1991, Golub1993, Ritchie1993}. SUD also encapsulates a Linux device
driver in a user space process, and similar to \sname, uses the IOMMU
and UML to achieve this~\cite{Boyd2010}. Microdriver is another solution,
which splits the driver between the kernel and user
space~\cite{Ganapathy2008}. It keeps the critical path code of the driver in
the kernel, but redirects the execution to the user space elsewhere.
Hunt~\cite{Hunt1997} also presents a solution that moves the driver to
the user space by employing a proxy in the kernel. LeVasseur et
al.~\cite{Levasseur2004}, on the other hand, execute the device driver
in a VM.
All these solutions improve the security of the operating system kernel
by removing the driver. However, in contrast to a \name, they
cannot improve the isolation between processes using the device, since
the driver is fully shared between them. Moreover, unlike a \name,
these solutions are reported to hurt the performance.

CuriOS~\cite{David2008} improves
the isolation between applications that use a microkernel service, such as
the file system, by providing the service with access to client states
only when the service is processing a request. This allows for
recovery from most of the errors in the service. In contrast, library
drivers improve the isolation between application by reducing the TCB
size.

Other solutions try to protect against errors in the drivers, either using
runtime and hardware-enforced techniques such as Nooks~\cite{Swift2003} or using
language-based techniques such as SafeDrive~\cite{Zhou2006}. In
contrast, a \name improves the system security by reducing the TCB size
through the principle of untrusted resource management.

Zhou et al.
move the device driver to the user space in order to create a trusted
path between the application and the device~\cite{Zhou2012, Zhou2014}.
However, they assume a single application
using an I/O device and therefore they assign the device to the
application. Unlike them, we tackle a more challenging problem where the
I/O device is shared between untrusting applications.

Schaelicke~\cite{Schaelicke2001} and Pratt~\cite{Pratt1997,
Pratt2001} propose hardware architectures to support secure user space
access to devices. In this work, we show that adequate hardware
primitives already exist on commodity accelerators, such as GPUs,
platforms to
run the device management code in the user space.

Exokernel~\cite{Engler1995}, U-Net~\cite{Von1995}, and
sv3~\cite{Stecklina2014} move part or all of the network device driver to the
user space for better security or performance. MyCloud SEP
detangles the resource management functions for disks and make them
untrusted in order to reduce
the TCB of virtualization platforms~\cite{Li2014}. We share the same goals
with this line of work. However, in our work, we demonstrate that such
an approach is applicable to a wider range of devices, especially
accelerators such as GPUs, and present a framework to apply similar
concepts to other devices as well.

Some existing device drivers incorporate a user space components in
addition to the kernel driver. Such drivers differ from library drivers
in one of two ways. First, the user space component is in the TCB since
it is shared by applications using the device. By compromising this user
space component, a malicious application can attack other applications.
This component is either a process (e.g., the driver host process in
Windows User-Mode Driver Framework (UMDF)~\cite{UMDF}) or a library loaded by
a shared process (e.g., Mali GPU drivers for Android, where libraries
are loaded by the surface\_flinger process and libusb, where the loader
process can control the device). Second, although the I/O stack
includes untrusted libraries, the kernel driver implements high-level
resource management with several APIs. For example, the in-kernel Display Miniport
Driver of Windows Display Driver Model (WDDM) implements \texttildelow70 APIs
including APIs for command submission~\cite{WDDM}. Mali kernel drivers for Linux
exports \texttildelow40 APIs. 

Some devices provide per-process hardware support. Examples are some NVIDIA
GPUs and Infiniband devices, which support per-process command queues.
While such hardware support is added for performance, it improves the
system security as well by reducing the TCB size, similar to library drivers.
Our work shows that similar goals can be achieved for devices without
such explicit hardware support as well.

\subsubsection*{Virtualization}

gVirt~\cite{Tian2014} supports mediated passthrough of Intel GPUs.
Privileged modules running in the Xen hypervisor and Dom0 give VMs
direct access to performance-critical resources of the GPU while
emulating their access to sensitive resources, allowing VMs to use the
native device driver. In contrast, \names are designed to securely
multiplex a device between processes in the same operating system.
However, the hardware isolation techniques used in gVirt can be
leveraged in \names for Intel GPUs as well.

The nonkernel~\cite{Ben-Yehuda2013} is proposed to give applications
direct access to devices.
In contrast, we demonstrated that a device can be
shared between multiple applications if its resources are multiplexed at
a finer granularity.

Dune~\cite{Belay12} gives applications direct access to virtualization
hardware extensions. \Names benefit from virtualization extensions as
well. For example, the \core\ uses the IOMMU to isolate device DMA
targets. However, in contrast to Dune, the hardware features are used
by the \core\ running in the kernel.

Paradice~\cite{AmiriSani2014} paravirtualizes I/O devices at the device
file interface and hence allows virtual machines to directly talk to the
device driver. For security, it uses a trusted hypervisor to provide
fault and device data isolation between virtual machines assuming that
the device driver is compromised. However, it cannot provide other
guarantees such as functional correctness. Library drivers are
complementary as they reduce the device driver TCB size and attack
surface, reducing the possibility of the device driver getting compromised
in the first place.

\subsubsection*{Other Accelerator Architectures}

Heterogeneous System Architecture~\cite{HSA} is a standard for future
accelerators, targeting both the accelerator hardware and software. One
hardware features of the HSA is the support for user space dispatch, allowing applications to
dispatch instructions to the accelerator without communicating with the driver
in the kernel. In our work, we demonstrated that such a feature is
feasible even with commodity accelerators. Moreover, we anticipate that
other resource management tasks, such as memory management, still remain
in the trusted device drivers for HSA-compliant devices, whereas a \name
makes all the management code untrusted.

\section{Discussions}

\subsection{Pros and Cons of Library Drivers}
\label{sec:pros_cons}

Other than improved security and performance, 
\names have three other advantages and three
disadvantages. The advantages are as follows.
First, \names allow each application to customize its own \lib.
For example, applications can trade-off the initial cost of allocating
a pool of memory buffers with the runtime cost of allocating buffers as
needed.
Second, \names greatly simplify driver development because the
developer can use user space debugging frameworks or high-level
languages, none of them are available in the kernel. Moreover,
developers can use a unified debugging framework for both the
application and the \lib, which can greatly help with
timing bugs and data races.
Third, \names improve memory usage accounting by the operating system.
\Old drivers for some devices, such as GPUs, implement their own memory
management, which gives applications a side channel to allocate parts of
the system memory, invisible to the operating system kernel for
accounting. In contrast, with a \name, all the system memory allocated
by a \lib is through the standard operating system memory management
interface, e.g., the {\tt mmap} and {\tt brk} syscalls in Linux.

The \name design has the following disadvantages.
First, \names complicate multi-process programming. For
example, sharing memory buffers between processes is easily done in
a \old driver, but requires 
peer-to-peer communication between the \libs in a \name.
Second, \names incurs launch time overheads to applications.
We evaluated this overhead in~\S\ref{sec:eval_overheads} and showed that
it is not significant.
Third, depending on the device, a \name may achieve coarser-grained
device memory sharing for applications. A \old driver can share the
device memory at the granularity of buffers, whereas with devices
without paging support for their own memory, such as the Radeon GPU in
our setup, device memory can only be shared between \libs using
contiguous segments.

\subsection{Devices with Multiple Hardware Execution Contexts}
\label{sec:multi_context}

For devices with multiple execution contexts,
the \core\ should bind and revoke the hardware contexts to \libs
independently. This puts new requirements on the device and platform
hardware as we will explain next.

First, the hardware management interface for different contexts must be
non-overlapping and isolated. That is, registers for each context should be
separate and instructions should only affect the resources of the given
context.
Also, device
DMA requests must be attributable to different contexts at a low level
so that an IOMMU can be used to isolate them.

As an example, self-virtualized devices~\cite{Raj2007, Willmann2007,
nvidia_white_paper, Dong2008} export multiple virtual devices, each of
which can be separately assigned to a VM. As a result, they readily
provide all the hardware primitives for the \core\ to bind different
virtual devices to different \libs.

\subsection{Radeon Instruction Set}

We allow a \lib to directly dispatch instructions to our Radeon and
Intel GPUs since the instructions are not sensitive. We noticed
a curious case with the Radeon GPU's instruction set though. Using the
instructions, an application can write to the registers of GPU
processing units in order to program and use them. Fortunately, it is
not possible to use the instructions to write to registers of components
of the GPU that affect the isolation, such as the memory controller.
However, we noticed that the Linux open source Radeon driver returns
error when inspecting the instructions submitted by an applications and
detecting accesses to large register numbers, which surprisingly
correspond to no actual registers according to the AMD reference
guide~\cite{evergreen_3d_registers} and the register layout in the
driver itself. Therefore, we believe that this is a simple correctness
check by the driver and not a security concern. Therefore, we currently
do not employ such a check in \core, although that is a possibility.
Adding the check to the \core will increase the TCB size by about a few
kLoC, but should not degrade the performance compared to what was
reported in this paper since we already performed these checks in the
\lib in our benchmarks (although it was not necessary).

\section{Conclusions}

We presented \names, a driver design that improves system security by
running device management code in untrusted libraries within application
processes rather than in the kernel. We discussed the device and
platform hardware properties needed for a \name and showed that they are
mostly available for commodity accelerators, such as GPUs, which are of
interest to us. We presented \sname, a \name implementation for two GPUs
of popular brands, Radeon and Intel. Our evaluation showed that \sname
reduced the size and attack surface of the TCB. Moreover, it improved
the performance for benchmarks requiring intensive interactions with the
driver. We believe that \names are a viable solution for accelerators,
an increasing important subset of devices.

\bibliographystyle{plain}
\bibliography{iovirt}

\begin{thebibliography}{10}

\bibitem{WDDM}
{Functions Registered by DriverEntry of Display Miniport Driver in Windows
  Display Driver Model (WDDM)}.
\newblock
  \url{http://msdn.microsoft.com/en-us/library/windows/hardware/ff566463(v=vs.%
85).aspx}.

\bibitem{GalliumCompute}
{GalliumCompute}.
\newblock \url{http://dri.freedesktop.org/wiki/GalliumCompute/}.

\bibitem{BUG2}
{Integer overflow in Linux DRM driver (CVE-2012-0044)}.
\newblock \url{https://bugzilla.redhat.com/show_bug.cgi?id=772894}.

\bibitem{Mesa}
{Mesa}.
\newblock \url{http://www.mesa3d.org/}.

\bibitem{nvidia_white_paper}
{NVIDIA GRID K1 and K2 Graphics-Accelerated Virtual Desktops and Applications.
  NVIDIA White Paper}.

\bibitem{dl}
{OpenGL Microbenchmark: Display List}.
\newblock \url{http://www.songho.ca/opengl/gl_displaylist.html}.

\bibitem{va}
{OpenGL Microbenchmark: Vertex Array}.
\newblock \url{http://www.songho.ca/opengl/gl_vertexarray.html}.

\bibitem{BUG5}
{Privilege escalation using an exploit in Linux NVIDIA binary driver}.
\newblock \url{http://seclists.org/fulldisclosure/2012/Aug/4}.

\bibitem{BUG1}
{Privilege escalation using NVIDIA GPU driver bug (CVE-2012-4225)}.
\newblock \url{http://www.securelist.com/en/advisories/50085}.

\bibitem{BUG4}
{Stack buffer overflow in NVIDIA display driver service in Windows 7}.
\newblock
  \url{https://www.securityweek.com/researcher-unwraps-dangerous-nvidia-driver%
-exploit-} \url{christmas-day}.

\bibitem{bromium_white_paper}
{Understanding Bromium Micro-virtualization for Security Architects. Bromium
  White Paper}.

\bibitem{BUG3}
{Unprivileged GPU access vulnerability using NVIDIA driver bug
  (CVE-2013-5987)}.
\newblock
  \url{http://nvidia.custhelp.com/app/answers/detail/a_id/3377/~/unprivileged-%
gpu-access-vulnerability---cve-2013-5987}.

\bibitem{VMDq}
{VMDq}.
\newblock
  \url{http://www.intel.com/content/www/us/en/network-adapters/gigabit-network%
-adapters/io-acceleration-technology-vmdq.html}.

\bibitem{WebGL_Security}
{WebGL Security}.
\newblock \url{http://www.khronos.org/webgl/security/}.

\bibitem{UMDF}
{Windows User-Mode Driver Framework (UMDF)}.
\newblock
  \url{http://msdn.microsoft.com/en-us/library/windows/hardware/ff560442(v=vs.%
85).aspx}.

\bibitem{Abramson2006}
D.~Abramson, J.~Jackson, S.~Muthrasanallur, G.~Neiger, G.~Regnier, R.~Sankaran,
  I.~Schoinas, R.~Uhlig, B.~Vembu, and J.~Wiegert.
\newblock {Intel Virtualization Technology for Directed I/O}.
\newblock {\em Intel Technology Journal}, 2006.

\bibitem{AmiriSani2014}
A.~Amiri~Sani, K.~Boos, S.~Qin, and L.~Zhong.
\newblock {I/O Paravirtualization at the Device File Boundary}.
\newblock In {\em Proc. ACM ASPLOS}, 2014.

\bibitem{Belay12}
A.~Belay, A.~Bittau, A.~Mashtizadeh, D.~Terei, D.~Mazieres, and C.~Kozyrakis.
\newblock {Dune: Safe User-level Access to Privileged CPU Features}.
\newblock In {\em Proc. USENIX OSDI}, 2012.

\bibitem{Ben-Yehuda2010}
M.~Ben-Yehuda, M.~D. Day, Z.~Dubitzky, M.~Factor, N.~Har'El, A.~Gordon,
  A.~Liguori, O.~Wasserman, and B.~A. Yassour.
\newblock {The Turtles Project: Design and Implementation of Nested
  Virtualization}.
\newblock In {\em Proc. USENIX OSDI}, 2010.

\bibitem{Ben-Yehuda2013}
M.~Ben-Yehuda, O.~Peleg, O.~Agmon Ben-Yehuda, I.~Smolyar, and D.~Tsafrir.
\newblock {The nonkernel: A Kernel Designed for the Cloud}.
\newblock In {\em Proc. ACM APSYS}, 2013.

\bibitem{Boyd2010}
S.~Boyd-Wickizer and N.~Zeldovich.
\newblock {Tolerating malicious device drivers in Linux}.
\newblock In {\em Proc. USENIX ATC}, 2010.

\bibitem{David2008}
F.~M. David, E.~Chan, J.~C. Carlyle, and R.~H. Campbell.
\newblock {CuriOS: Improving Reliability through Operating System Structure}.
\newblock In {\em Proc. USENIX OSDI}, 2008.

\bibitem{Dike2001}
J.~Dike.
\newblock {User-Mode Linux}.
\newblock In {\em Proc. Ottawa Linux Symposium}, 2001.

\bibitem{Dong2008}
Y.~Dong, Z.~Yu, and G.~Rose.
\newblock {SR-IOV Networking in Xen: Architecture, Design and Implementation}.
\newblock In {\em Proc. USENIX Workshop on I/O Virtualization (WIOV)}, 2008.

\bibitem{Douceur2008}
J.~R. Douceur, J.~Elson, J.~Howell, and J.~R. Lorch.
\newblock {Leveraging Legacy Code to Deploy Desktop Applications on the Web.}
\newblock In {\em Proc. USENIX OSDI}, 2008.

\bibitem{Elphinstone2013}
K.~Elphinstone and G.~Heiser.
\newblock {From L3 to seL4 What Have We Learnt in 20 Years of L4 Microkernels?}
\newblock In {\em Proc. ACM SOSP}, 2013.

\bibitem{Engler1995}
D.~R. Engler, M.~F. Kaashoek, and J.~O'Toole Jr.
\newblock {Exokernel: an Operating System Architecture for Application-Level
  Resource Management}.
\newblock In {\em Proc. ACM SOSP}, 1995.

\bibitem{Forin1991}
A.~Forin, D.~Golub, and B.~N. Bershad.
\newblock {An I/O System for Mach 3.0}.
\newblock In {\em Proc. USENIX Mach Symposium}, 1991.

\bibitem{Ganapathi2006}
A.~Ganapathi, V.~Ganapathi, and D.~Patterson.
\newblock {Windows XP Kernel Crash Analysis}.
\newblock In {\em Proc. USENIX LISA}, 2006.

\bibitem{Ganapathy2008}
V.~Ganapathy, M.~J. Renzelmann, A.~Balakrishnan, M.~M. Swift, and S.~Jha.
\newblock {The Design and Implementation of Microdrivers}.
\newblock In {\em Proc. ACM ASPLOS}, 2008.

\bibitem{Golub1993}
David~B. Golub, Guy~G. Sotomayor, and Freeman~L. Rawson, III.
\newblock {An Architecture for Device Drivers Executing As User-Level Tasks}.
\newblock In {\em Proc. USENIX MACH III Symposium}, 1993.

\bibitem{Gordon2012}
A.~Gordon, N.~Amit, N.~Har'El, M.~Ben-Yehuda, A.~Landau, D.~Tsafrir, and
  A.~Schuster.
\newblock {ELI: Bare-Metal Performance for I/O Virtualization}.
\newblock In {\em Proc. ACM ASPLOS}, 2012.

\bibitem{Gupta2011}
V.~Gupta, K.~Schwan, N.~Tolia, V.~Talwar, and P.~Ranganathan.
\newblock {Pegasus: Coordinated Scheduling for Virtualized Accelerator-Based
  Systems}.
\newblock In {\em USENIX ATC}, 2011.

\bibitem{Howell2013}
J.~Howell, B.~Parno, and J.~Douceur.
\newblock {Embassies: Radically Refactoring the Web}.
\newblock In {\em Proc. USENIX NSDI}, 2013.

\bibitem{Hunt1997}
G.~C. Hunt.
\newblock {Creating User-Mode Device Drivers with a Proxy}.
\newblock In {\em Proc. USENIX Windows NT Workshop}, 1997.

\bibitem{evergreen_3d_registers}
Advanced Micro~Devices Inc.
\newblock {Radeon Evergreen 3D Register Reference Guide, Revision 1.0}.
\newblock \url{http://www.x.org/docs/AMD/old/evergreen_3D_registers_v2.pdf},
  2011.

\bibitem{TI_keystone}
Texas Instruments.
\newblock {KeyStone Architecture Multicore Shared Memory Controller (MSMC) -
  User Guide}.
\newblock Literature Number: SPRUGW7A, 2011.

\bibitem{Kato2011}
S.~Kato, K.~Lakshmanan, R.~R. Rajkumar, and Y.~Ishikawa.
\newblock {TimeGraph: GPU Scheduling for Real-time Multi-tasking Environments}.
\newblock In {\em Proc. USENIX ATC}, 2011.

\bibitem{nvidia_hw_doc}
M.~Ko\'{s}cielnicki.
\newblock {NVIDIA Hardware Documentation}.
\newblock
  \url{https://media.readthedocs.org/pdf/envytools/latest/envytools.pdf}.

\bibitem{HSA}
George Kyriazis.
\newblock {Heterogeneous System Architecture: A Technical Review, AMD White
  Paper}.
\newblock 2012.

\bibitem{Leslie2005}
B.~Leslie, P.~Chubb, N.~Fitzroy-Dale, S.~G{\"o}tz, C.~Gray, L.~Macpherson,
  D.~Potts, Y.~Shen, K.~Elphinstone, and G.~Heiser.
\newblock {User-Level Device Drivers: Achieved Performance}.
\newblock {\em Journal of Computer Science and Technology}, 20(5), 2005.

\bibitem{Levasseur2004}
J.~LeVasseur, V.~Uhlig, J.~Stoess, and S.~G{\"o}tz.
\newblock {Unmodified Device Driver Reuse and Improved System Dependability via
  Virtual Machines}.
\newblock In {\em Proc. USENIX OSDI}, 2004.

\bibitem{Li2014}
M.~Li, Z.~Zha, W.~Zang, M.~Yu, P.~Liu, and K.~Bai.
\newblock {Detangling Resource Management Functions from the TCB in
  Privacy-Preserving Virtualization}.
\newblock In {\em Proc. European Symposium on Research in Computer Security
  (ESORICS)}, 2014.

\bibitem{Liu2006}
J.~Liu, W.~Huang, B.~Abali, and D.~K. Panda.
\newblock {High Performance VMM-Bypass I/O in Virtual Machines}.
\newblock In {\em Proc. USENIX ATC}, 2006.

\bibitem{Cortex_A15_TRM}
ARM Ltd.
\newblock {ARM Cortex-A15 Technical Reference Manual}.
\newblock {\em ARM DDI}, 0438C, 2011.

\bibitem{Menon2012}
J.~Menon, M.~De~Kruijf, and K.~Sankaralingam.
\newblock {iGPU: Exception Support and Speculative Execution on GPUs}.
\newblock In {\em Proc. IEEE ISCA}, 2012.

\bibitem{Menychtas2014}
K.~Menychtas, K.~Shen, and M.~L. Scott.
\newblock {Disengaged Scheduling for Fair, Protected Access to Fast
  Computational Accelerators}.
\newblock In {\em Proc. ACM ASPLOS}, 2014.

\bibitem{Menychtas2013}
Konstantinos Menychtas, Kai Shen, and Michael~L Scott.
\newblock {Enabling OS Research by Inferring Interactions in the Black-Box GPU
  Stack}.
\newblock In {\em Proc. USENIX ATC}, 2013.

\bibitem{Pichai2014}
B.~Pichai, L.~Hsu, and A.~Bhattacharjee.
\newblock {Architectural Support for Address Translation on GPUs}.
\newblock In {\em Proc. ACM ASPLOS}, 2014.

\bibitem{Popek1974}
G.~J. Popek and R.~P. Goldberg.
\newblock {Formal requirements for virtualizable third generation
  architectures}.
\newblock {\em Communications of the ACM}, 1974.

\bibitem{Porter2011}
D.~E. Porter, S.~Boyd-Wickizer, J.~Howell, R.~Olinsky, and G.~C. Hunt.
\newblock {Rethinking the Library OS from the Top Down}.
\newblock In {\em Proc. ACM ASPLOS}, 2011.

\bibitem{Pratt2001}
I.~Pratt and K.~Fraser.
\newblock {Arsenic: A User-Accessible Gigabit Ethernet Interface}.
\newblock In {\em Proc. IEEE INFOCOM}, 2001.

\bibitem{Pratt1997}
Ian~A. Pratt.
\newblock {The User-Safe Device I/O Architecture}.
\newblock {\em Doctoral thesis, University of Cambridge}, 1997.

\bibitem{Raj2007}
H.~Raj and K.~Schwan.
\newblock {High Performance and Scalable I/O Virtualization via
  Self-Virtualized Devices}.
\newblock In {\em Proc. ACM HPDC}, 2007.

\bibitem{Ritchie1993}
D.~S. Ritchie and G.~W. Neufeld.
\newblock {User Level IPC and Device Management in the Raven Kernel}.
\newblock In {\em USENIX Microkernels and Other Kernel Architectures
  Symposium}, 1993.

\bibitem{Schaelicke2001}
L.~Schaelicke.
\newblock {Architectural Support for User-Level I/O}.
\newblock {\em Doctoral thesis, University of Utah}, 2001.

\bibitem{Soares2010}
L.~Soares and M.~Stumm.
\newblock {FlexSC: Flexible System Call Scheduling with Exception-Less System
  Calls}.
\newblock In {\em Proc. USENIX OSDI}, 2010.

\bibitem{Stecklina2014}
J.~Stecklina.
\newblock {Shrinking the Hypervisor One Subsystem at a Time: A Userspace Packet
  Switch for Virtual Machines}.
\newblock In {\em Proc. ACM VEE}, 2014.

\bibitem{Swift2003}
M.~M. Swift, B.~N. Bershad, and H.~M. Levy.
\newblock {Improving the Reliability of Commodity Operating Systems}.
\newblock In {\em Proc. ACM SOSP}, 2003.

\bibitem{Tian2014}
K.~Tian, Y.~Dong, and D.~Cowperthwaite.
\newblock {A Full GPU Virtualization Solution with Mediated Pass-Through}.
\newblock In {\em Proc. USENIX ATC}, 2014.

\bibitem{Von1995}
T.~Von~Eicken, A.~Basu, V.~Buch, and W.~Vogels.
\newblock {U-Net: A User-Level Network Interface for Parallel and Distributed
  Computing}.
\newblock In {\em Proc. ACM SOSP}, 1995.

\bibitem{Waldspurger2002}
C.~A. Waldspurger.
\newblock {Memory Resource Management in VMware ESX Server}.
\newblock {\em ACM SIGOPS Operating Systems Review}, 2002.

\bibitem{Willmann2007}
P.~Willmann, J.~Shafer, D.~Carr, A.~Menon, S.~Rixner, A.~L. Cox, and
  W.~Zwaenepoel.
\newblock {Concurrent Direct Network Access for Virtual Machine Monitors}.
\newblock In {\em Proc. IEEE High Performance Computer Architecture (HPCA)},
  2007.

\bibitem{Zhou2006}
F.~Zhou, J.~Condit, Z.~Anderson, I.~Bagrak, R.~Ennals, M.~Harren, G.~Necula,
  and E.~Brewer.
\newblock {SafeDrive: Safe and Recoverable Extensions Using Language-Based
  Techniques}.
\newblock In {\em Proc. USENIX OSDI}, 2006.

\bibitem{Zhou2012}
Z.~Zhou, V.~D. Gligor, J.~Newsome, and J.~M. McCune.
\newblock {Building Verifiable Trusted Path on Commodity x86 Computers}.
\newblock In {\em Proc. IEEE Symposium on Security and Privacy (S\&P)}, 2012.

\bibitem{Zhou2014}
Z.~Zhou, M.~Yu, and V.~D. Gligor.
\newblock {Dancing with Giants: Wimpy Kernels for On-demand Isolated I/O}.
\newblock In {\em Proc. IEEE Symposium on Security and Privacy (S\&P)}, 2014.

\end{thebibliography}

\end{document}